\documentclass[12pt]{iopart}
\usepackage{graphicx}

\begin{document}

\title[]{The freeze-out properties of hyperons in a microscopic transport model}

\author{Zhenglian Xie$^{1,2}$, Pingzhi Ning$^{1}$ and Steffen A. Bass$^2$}

\address{$^1$ School of Physics, Nankai University,
             Tianjin 300071, China}
\address{$^2$ Department of Physics, Duke University,
             Durham, North Carolina 27708-0305, USA}

\ead{xiezhenglian@mail.nankai.edu.cn}

\begin{abstract}
The excitation function of freeze-out time, average freeze-out temperature and freeze-out energy
density of (multi-) strange baryons created in relativistic heavy-ion collisions
is investigated in the framework of a microscopic transport model. We find that the
$\Omega$ on average freezes out
earlier than the nucleon, $\Lambda$ and $\Xi$.
 The average freeze-out temperature and energy
density as well as the spread between the different baryonic states
increase monotonously with increasing beam energy and should approach
a universal value in the case of a hadronizing Quark-Gluon-Plasma.
\end{abstract}

\submitto{\JPG}
\maketitle

\section{Introduction}

A major goal of colliding heavy-ions at relativistic energies is to heat up a
small region of space-time to temperatures as high as are thought
to have occurred during the
early evolution of the universe, a few microseconds after the big
bang~\cite{Kolb:1990vq}. In ultra-relativistic heavy-ion collisions, such
as are currently being explored at the Relativistic Heavy-Ion Collider (RHIC),
the four-volume of hot and dense matter,
with temperatures  above $\sim150$~MeV, is
on the order of $\sim (10$~fm$)^4$. The state of strongly
interacting matter at such high temperatures (or density of
quanta) is usually called quark-gluon plasma (QGP)~\cite{Collins:1974ky,Shuryak:1980tp}.

The first five years of RHIC operations
at  $\sqrt{s_{NN}}=130$~GeV and $\sqrt{s_{NN}}=200$~GeV
have yielded a vast amount of interesting and
sometimes surprising results \cite{Adcox:2004mh,Back:2004je,Adams:2005dq,Arsene:2004fa}, many of
which have not yet been fully evaluated or understood by theory.
There exists mounting evidence that RHIC has created
a hot and dense state of deconfined QCD matter with properties similar to
that of an ideal fluid \cite{Ludlam:2005gx,Gyulassy:2004zy} -- this state of matter
has been termed the {\em strongly interacting Quark-Gluon-Plasma} (sQGP).

The central problem in the study of the sQGP is that
the deconfined quanta of a sQGP are not directly
observable due to the fundamental confining property of the
physical QCD vacuum. If we could see free quarks and gluons
(as in ordinary plasmas) it would be trivial to verify the QCD
prediction of the QGP state. However, nature chooses to hide those
constituents within  the confines of color neutral composite many
body systems -- hadrons. One of the main tasks in relativistic
heavy-ion research is to find clear and unambiguous connections
between the transient (partonic) plasma state and the observable hadronic final
state (for reviews on QGP signatures, please see \cite{Harris:1996zx,Bass:1998vz}).
Among the currently pressing issues in the exploration of the (s)QGP
is the question at temperature and energy-density deconfinement is achieved.
In order to address this question, a low energy run is being performed
at RHIC and the new FAIR project at GSI is aiming in the same direction.
What is now needed most is a set of robust signatures which could indicate
the transition from an excited hadron gas to a deconfined state of 
(strongly interacting) quarks and gluons.

The investigation of strangeness production in relativistic heavy 
ion collisions has been proven to be a powerful tool for the study 
of highly excited nuclear matter, both in terms of the reaction
dynamics and in terms of its hadrochemistry
\cite{Rafelski:1982pu,Koch:1986ud,Koch:1988nn,BraunMunzinger:1994xr,Letessier:1996ad,Vance:1999pr,Soff:1999et,BraunMunzinger:1999qy,Rafelski:2001hp,BraunMunzinger:2001ip}.
Furthermore, strangeness has  been suggested as a signature
for the creation of a Quark-Gluon-Plasma (QGP) 
\cite{Rafelski:1982pu,Koch:1986ud,Koch:1988nn}:
in the final state of a heavy-ion collision, subsequent to the formation
and decay of a QGP, strangeness has been predicted to
be enhanced relative to the strangeness yield in
elementary hadron+hadron collisions.

The slope parameters of (multi-)strange baryons transverse momentum distributions 
have been measured at the CERN
SPS \cite{Andersen:1998vu,Appelshauser:1998va,Bearden:1996dd} and at 
RHIC \cite{Adams:2003fy,Abelev:2007xp}. 
They exhibit a characteristic deviation from a hydrodynamically
motivated blastwave expansion \cite{vanHecke:1998yu}. This {\em reduction} in the slope
parameter has been shown in microscopic and hybrid macro+micro transport
models to be caused early freeze-out of (multi-)strange
baryons due to their reduced hadronic cross sections, compared to non-strange
baryons \cite{vanHecke:1998yu,Dumitru:1999sf,Nonaka:2006yn,Hirano:2007ei}. 
It is the purpose of our analysis to explore whether this
feature is prevalent at lower beam energies as well and whether it can possibly
be exploited as an indicator for the transition from an excited hadron gas
to a (s)QGP.

\section{The UrQMD model}
\label{UrQMD_section}
For our studies we use the Ultra-relativistic Quantum Molecular Dynamics
(UrQMD) model,
a microscopic hadronic transport model with hadron and constituent
(di-)quark degrees of freedom.
In UrQMD system evolves as a sequence of binary
collisions or $2-N$-body decays  of mesons, baryons, and constituent (di-)quarks.
Binary collisions are performed in a point-particle sense:
Two particles collide if their minimum distance $d$,
i.e.\ the minimum relative
distance of the centroids of the Gaussians during their motion,
in their CM frame fulfills the requirement:
\begin{equation}
 d \le d_0 = \sqrt{ \frac { \sigma_{\rm tot} } {\pi}  }  , \qquad
 \sigma_{\rm tot} = \sigma(\sqrt{s},\hbox{ type} ).
\end{equation}
The cross section is assumed to be the free cross section of the
regarded collision type ($N-N$, $N-\Delta$, $\pi-N$ \ldots).

Especially when studying the hadrochemistry of a heavy-ion reaction
it is of great importance to include as many hadronic states as possible
into the model calculation:
The UrQMD collision term contains 49 different baryon species
(including nucleon, delta and hyperon resonances with masses up to 2 GeV)
and 25 different meson species (including strange meson resonances), which
are supplemented by their corresponding anti-particle
and all isospin-projected states.
Full baryon/antibaryon symmetry is included.
For excitations with higher masses a string picture is used.
All states listed can be produced in string decays, s-channel
collisions or resonance decays.

Tabulated or parameterized experimental cross sections are used when
available, resonance absorption and scattering is handled via the
principle of detailed balance. If no experimental information is
available, the cross section is either  calculated via an OBE model
or via a modified additive quark model which takes basic phase space
properties into account. A detailed overview of the elementary cross
sections and string excitation scheme included in the UrQMD model is
given elsewhere \cite{Bass:1998ca,Bleicher:1999xi}.

\section{Strangeness production in UrQMD}

Strangeness may be produced either in initial collisions among
the incoming nucleons
of the two colliding nuclei or through secondary interactions among
produced particles, e.g. pions and nucleons or excited resonances.

For collision energies which lead to the creation of a hadron gas
at temperatures up to approx. 140 MeV, the dominant strangeness
production mechanism in UrQMD utilizes a resonance gas approach:
here strangeness production is a two step
process. Initially a heavy baryon resonance is excited, e.g. via
$p+p\to N + N^*_{1710}$, which subsequently decays via
$ N^*_{1710} \to Y+K^+$. Subsequently, the hyperon could rescatter
with a kaon to form a $\Xi^*$ resonance, which could decay into
a $\Xi + \pi$ final state to facilitate $\Xi$ production.
This approach allows for an easy
extension into the higher energy domain and may provide some
rudimentary guidance for unknown strangeness production cross
sections in secondary collisions,
e.g. $\pi+N\to N^*_{1710} \to Y+K^+$.
The aforementioned secondary interactions like pion-induced strangeness production
or flavor-exchange reactions are at least as important for the reaction
dynamics and final strangeness yield as the initial/primordial strangeness
production channels.  The hyperon resonances which are excited via this cross section
may either decay again into the $K^-+N$ channel, or to almost equal
probability decay into the $Y+\pi$ channel, thus transferring strangeness
in and out of baryonic degrees of freedom. At beam energies in the SIS and
low AGS domain this
$K^-+N \leftrightarrow Y+\pi $ exchange reaction is of particular importance.

At higher incident beam energies, particle production in general is dominated
by string excitation and fragmentation. Up to 70\% of the total strangeness
produced in a Pb+Pb collision at top CERN-SPS energies is produced in initial
highly energetic nucleon-nucleon interactions which lead to the excitation
and subsequent fragmentation of strings. On an elementary
hadron-hadron level, the parameters of the string fragmentation
are fitted to measured multiplicities and momentum distributions.

For our analysis we have calculated and analyzed the time-evolution
of central Au+Au collisions at 10.6, 20, 40, 80 and 160 GeV/nucleon.
Typically we have calculated between 15K and 50K collisions per incident
beam energy. Figure~1 shows the transverse momentum spectra of baryons produced in
central Au+Au collisions at  20, 80 and 160 GeV/nucleon incident beam energy and
serves as a model baseline which can directly be compared to the final state
accessible by the experiments.

\begin{center}
\begin{figure}
\centering
\includegraphics[width=0.9\textwidth, angle=0]{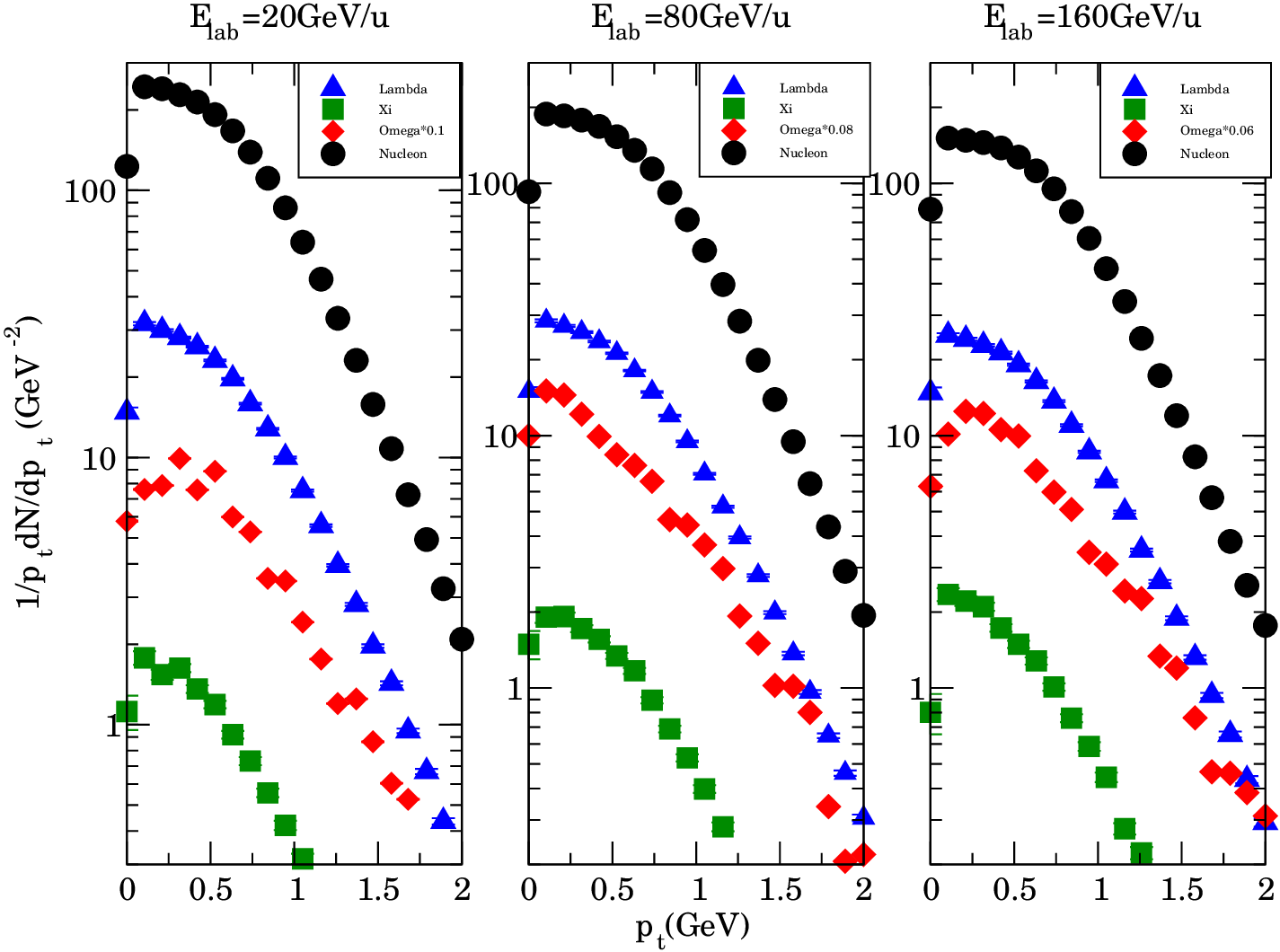}
 \caption{Transverse momentum spectra of baryons produced in
 central Au+Au collisions at mid-rapidity and 20, 80 and 160 GeV/nucleon incident beam energy.}
\end{figure}

\end{center}

\section{Methodology: freeze-out times, temperatures and energy-densities}


In this section we shall discuss the basic concepts which enter into
our analysis: the main task is to calculate the local temperature
and local energy-density a baryon experiences at any given time
during the collision evolution. We accomplish this by transforming the momenta
of all particles in the system into
the local rest-frame of the respective baryon, calculating the
kinetic energy of all surrounding hadrons within a radius of 1 fm, and
then utilizing the equipartition theorem to determine the local temperature.
The local energy density is trivially calculated by summing over the total
energy of all hadrons in the sphere and dividing by the volume of that sphere.
The radius of the sphere has been chosen such that it contains a sufficient
number hadrons for the temperature and energy-density extraction, but is still
sufficiently small to provide a local measure of temperature and energy-density.

The freeze-out time and position of a hadron is defined to be the location and
time of its last interaction, i.e. scattering. While the concept of a freeze-out
time is thus well-defined on the basis of an individual hadron, it is less
well-defined for a given particle species or even the full system as a whole.
Referring to a specific freeze-out time (or temperature) of  a hadron species, 
or even the entire collision system,
is only an approximation, which at most can
hold true on average. Figure~2 shows the distribution of freeze-out times at
mid-rapidity for the different baryon species in Au+Au collisions at incident
beam energies of 20 GeV/nucleon (left) and 160 GeV/nucleon (right). The
normalization is chosen such that all curves integrate to unity in order
to facilitate the comparison between the different hadron species.

\begin{center}
\begin{figure}
\includegraphics[width=0.47\textwidth, angle=0]{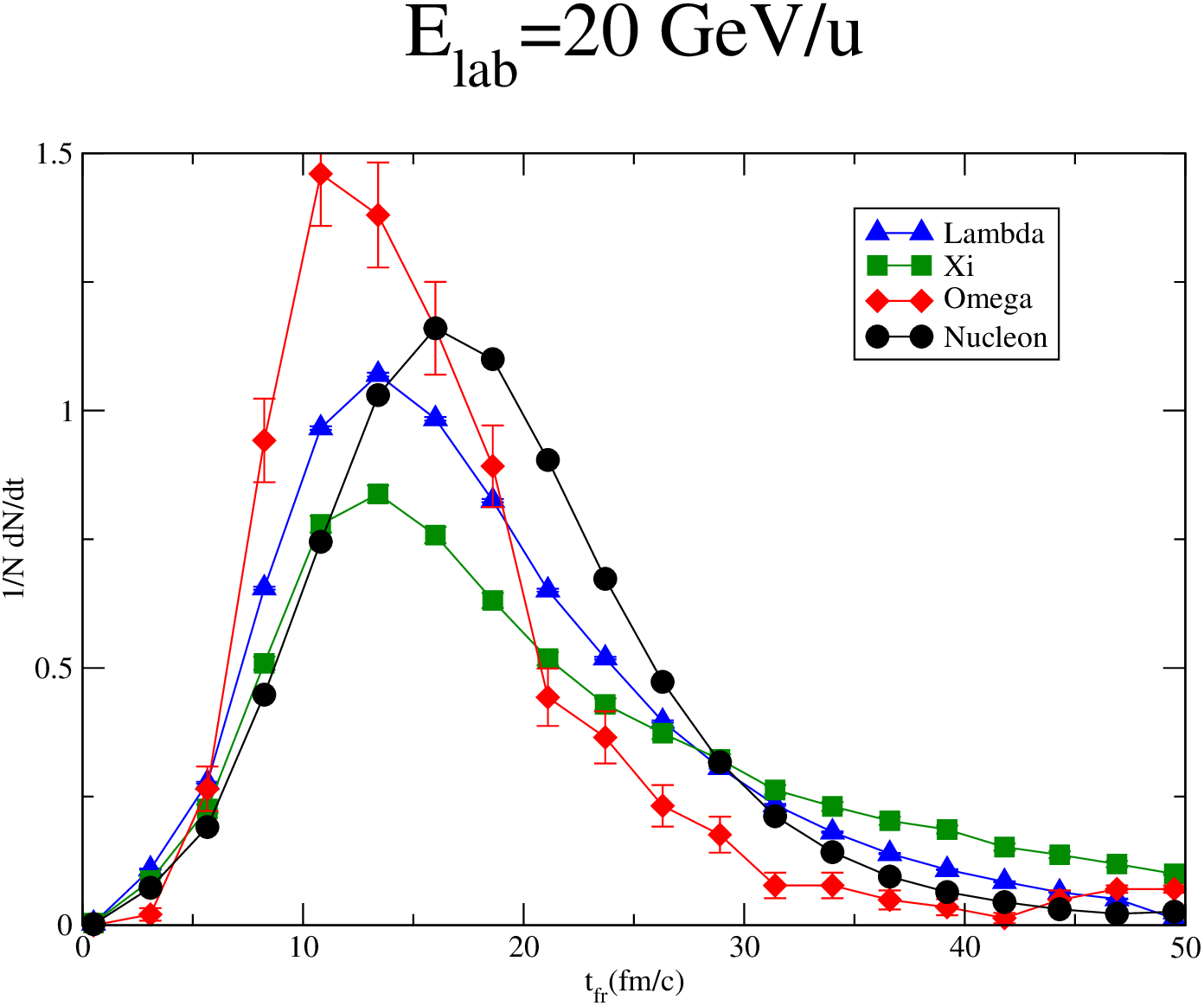}\hfill
\includegraphics[width=0.47\textwidth, angle=0]{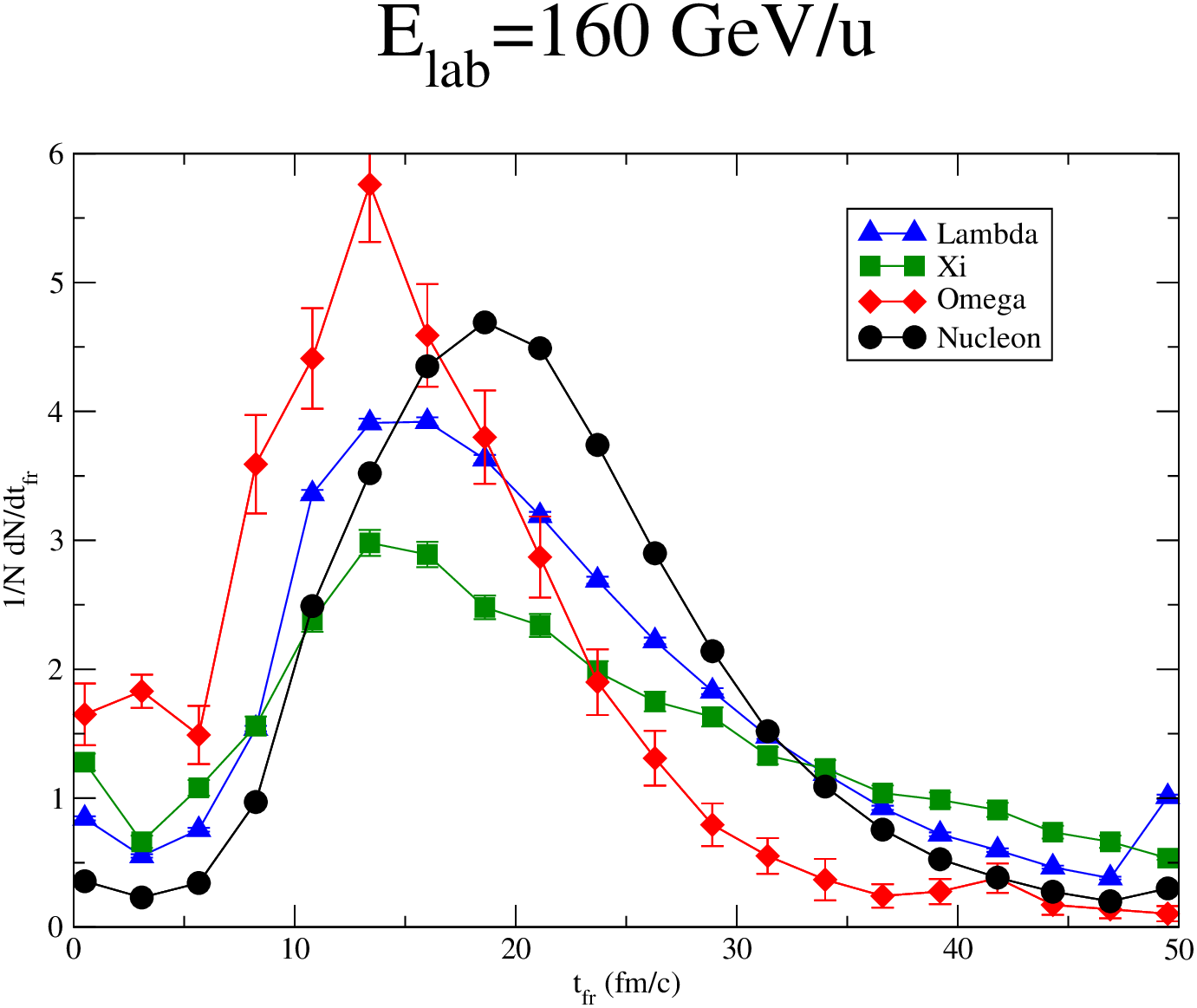}
 \caption{Normalized freeze-out time distributions for nucleons , Lambda's, Xi's and Omega's in Au+Au
 collisions at $E_{lab}$=20GeV/u (left) and $E_{lab}$=160 GeV/u (right).}
\end{figure}

\end{center}

As one can see in Figure~2, the freeze-out distributions for all
baryons are very broad - basically baryons are ejected from the
reaction zone continuously over the entire time-evolution
of the reaction. Nevertheless, some interesting trends are visible,
most notably the freeze-out time distribution of the Omega is peaked
significantly earlier than that of the other baryons. We would expect
the $\Xi$ to show a similar trend as the $\Omega$, but suffer from contamination
due to the long life-time (i.e. narrow decay width) of the $\Xi^*(1530)$ 
resonance, which is responsible for the long tail towards late times in
the freeze-out time distribution of the $\Xi$.
In order to facilitate our analysis of the Au+Au energy excitation
function, we need to condense the information present in these
freeze-out time distributions by averaging over these distributions.
This can be done in several ways: 
\begin{itemize}
\item the {\em average} freeze-out
time $\bar{t}_f$ of a species is being calculated by averaging
the individual freeze-out times of all particles of that species --
mathematically it constitutes the average of the distributions 
depicted in Figure~2. The shortcoming of this definition is that
the distributions are rather broad and that a significant fraction
of all particles of a given species may continue to interact
well beyond that time.
\item alternatively we can define the freeze-out time of a species
to be at that instant when e.g. 80\% of its particles have ceased
interacting: $t^{80\%}_f$. The advantage of this definition lies
in its better connection to the dynamics of the system, i.e.
most particles of the species have frozen out at that time.
\end{itemize}

\begin{table}
\caption{\label{label}The calculated freeze-out time of N,
$\Lambda$, $\Xi$ and $\Omega$.}
\begin{indented}
\item[]\begin{tabular}{@{}*{5}{l}}
\br
Energy(Gev/u)&N(fm/c)&$\Lambda(fm/c)$&$\Xi(fm/c)$&$\Omega(fm/c)$\\
\mr
10.6&18.8&18.9&24.1&16.7\\
20.0&18.8&18.9&24.6&16.1\\
40.0&19.2&19.3&24.2&15.8\\
80.0&20.1&20.2&24.7&15.8\\
160.0&21.2&21.2&25.5&16.2\\
\br
\end{tabular}
\end{indented}
\end{table}

\section{Results and discussion}

Let us first investigate the connection between the different
freeze-out time distributions of the final state baryon species
(nucleon, hyperon, Cascade and Omega) and their average freeze-out
temperature. We focus on baryons emitted within $\pm 1$ unit
around mid-rapidity. In Figure~3 we have calculated the average
freeze-out temperature of the four final state baryon species
as a function of time, i.e. at each time-step we have calculated
the average local temperature at the last previously encountered interaction point for
all nucleons, hyperons, cascades and omegas in the system. A leveling
off of that temperature signifies the freezing out of the majority
of baryons of that species and in the long time limit these curves
converge to the overall average freeze-out temperature for a given
species. The left frame of Figure~3 shows our analysis for an incident
beam energy of 20 GeV/nucleon and the right frame shows the same analysis
for 160 GeV/nucleon. The final average freeze-out temperature for 
N, $\Lambda$, $\Xi$ and $\Omega$ at $E_{lab}$=160GeV/nucleon are 
86.5, 115, 136 and 145 MeV respectively. A careful study of 
Figure~3 indicates at what time during the evolution of the
reaction the freeze-out temperature 'saturates' and the respective
baryon species thus becomes insensitive to the subsequent evolution
of the system. Comparing the two incident beam energies, we find
that there is very little difference in the freeze-out temperatures
among (multi-)strange baryon species at 20 GeV/nucleon and that
a more pronounced ordering according to strangeness content develops
at the higher beam energy. In both cases the nucleons exhibit a 
significantly lower freeze-out temperature than strange baryons.

\begin{center}
\begin{figure}
\includegraphics[width=0.47\textwidth, angle=0]{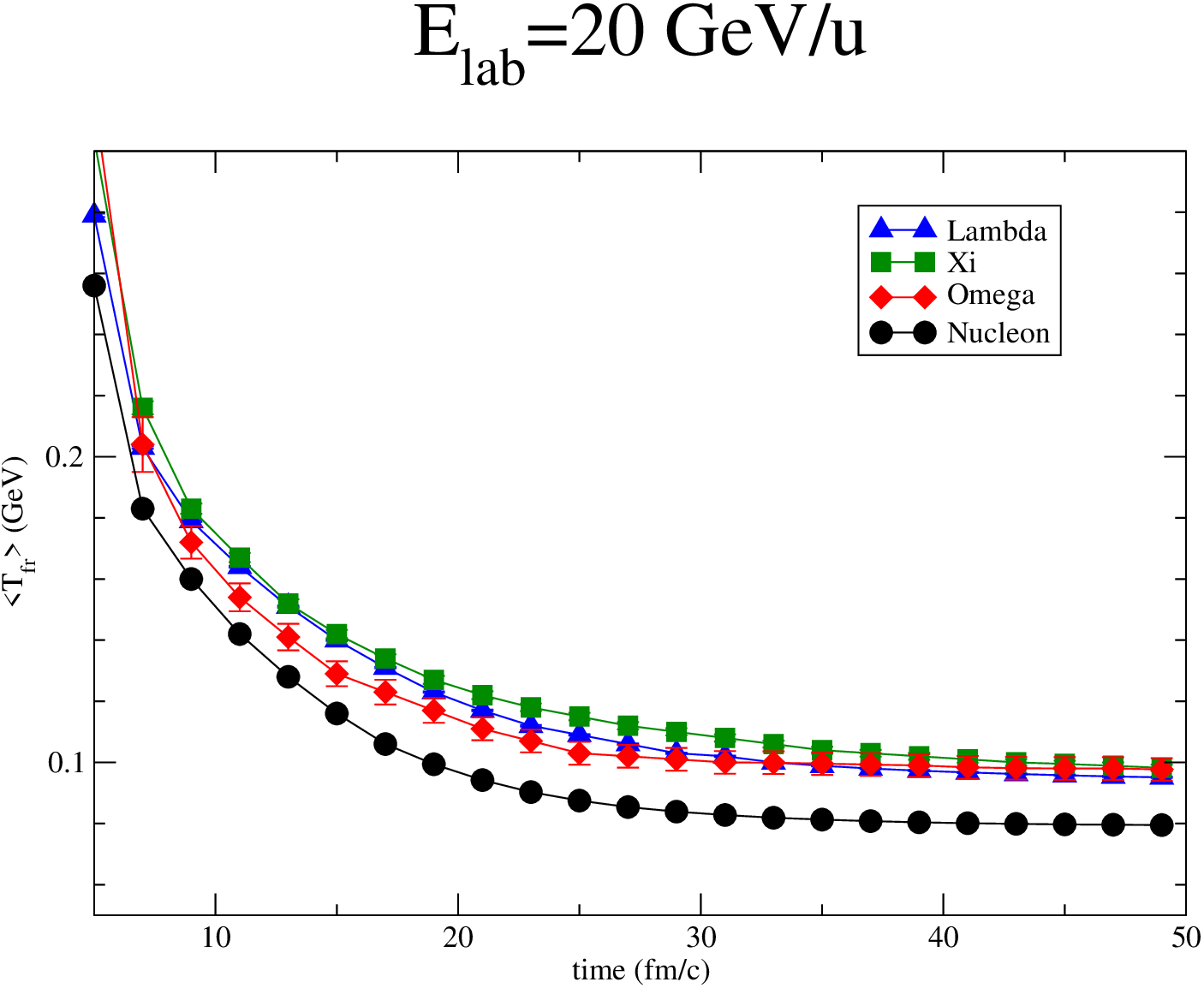}\hfill
\includegraphics[width=0.47\textwidth, angle=0]{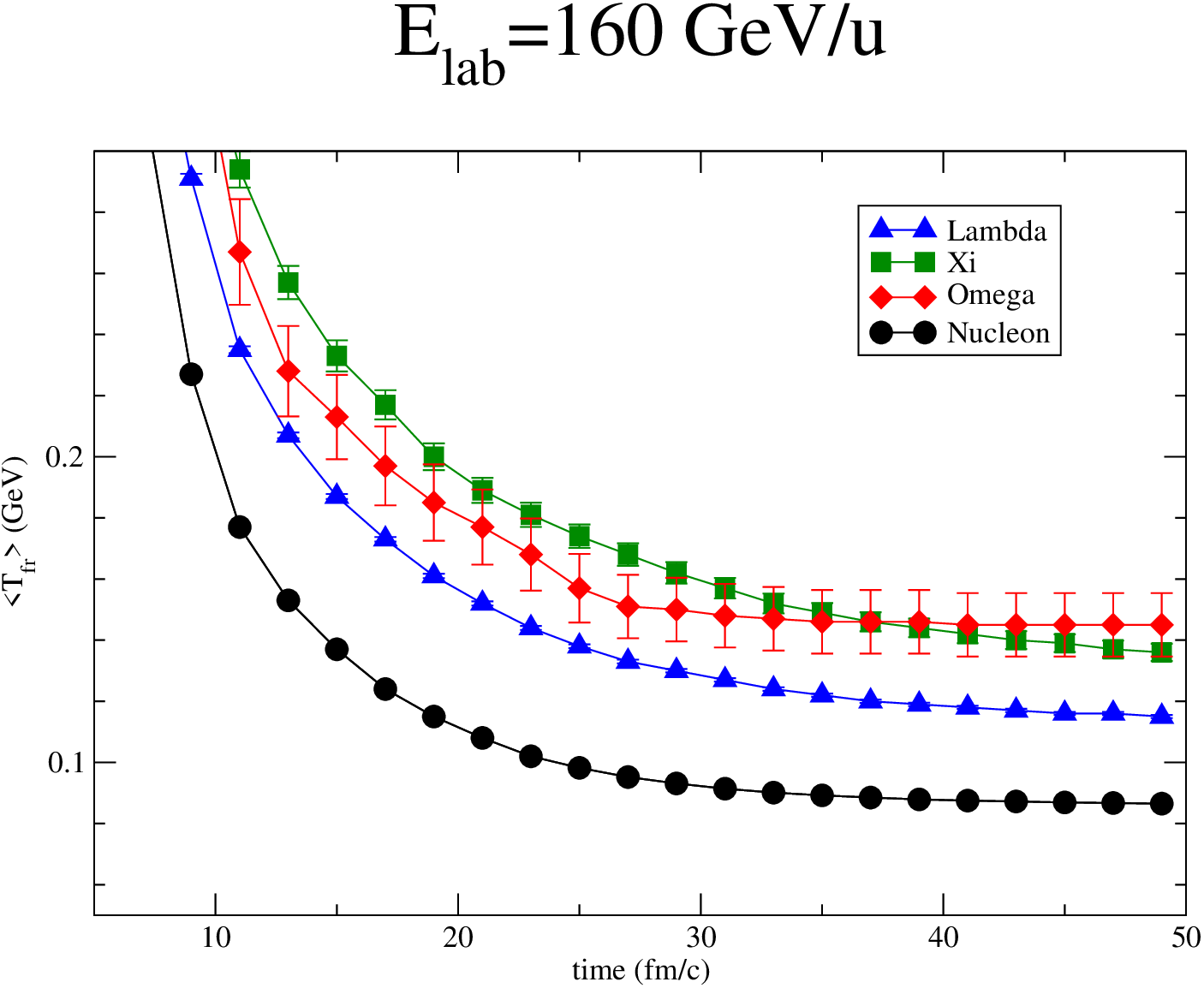}
 \caption{The average freeze-out temperature for Nucleons, Lambda's, Xi's and Omega's 
 as function of time for $E_{lab}$=20GeV/u (left) and $E_{lab}$=160GeV/u (right). A saturation
 of the respective curve indicates that the majority of particles of that species have ceased
 interacting.}
\end{figure}
\end{center}

Figure~4 shows the same analysis as Figure~3, but for the freeze-out
energy-density instead of the freeze-out temperature. It is interesting
to note that whereas at 160 GeV/nucleon we observe exactly the same trends
in terms of the freeze-out energy-density as we did for the freeze-out temperature,
at 20 GeV/nucleon the $\Omega$ seems to probe a higher freeze-out energy density
than the $\Xi$ and the hyperons, differing from our findings for the freeze-out temperature.
At 160 GeV/nucleon we find the average freeze-out energy
densities for N, $\Lambda$, $\Xi$ and $\Omega$ to be
0.3, 0.57, 0.85 and 0.84 GeV/$(fm)^{3}$ respectively.

\begin{center}
\begin{figure}
\includegraphics[width=0.47\textwidth, angle=0]{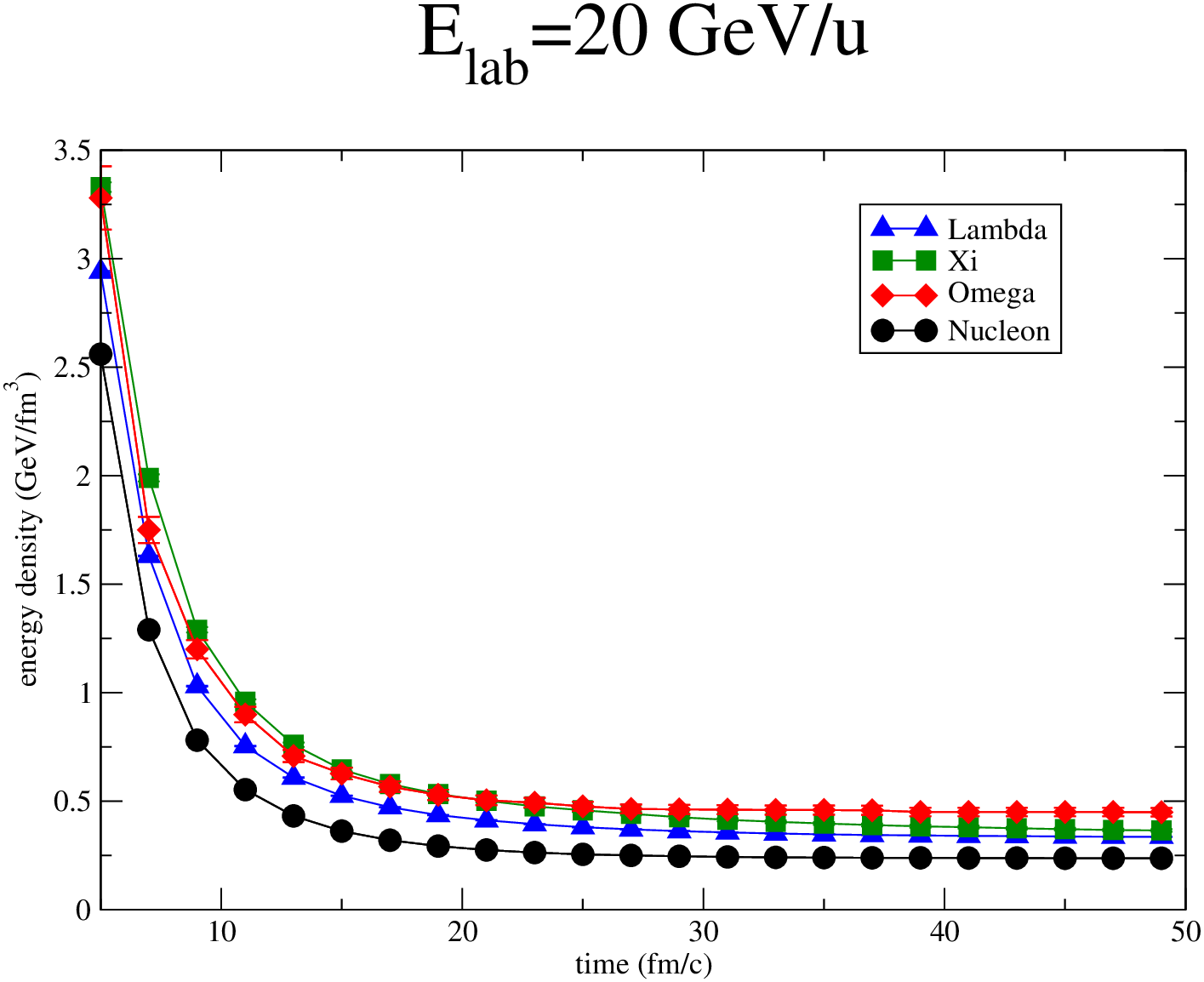}\hfill
\includegraphics[width=0.47\textwidth, angle=0]{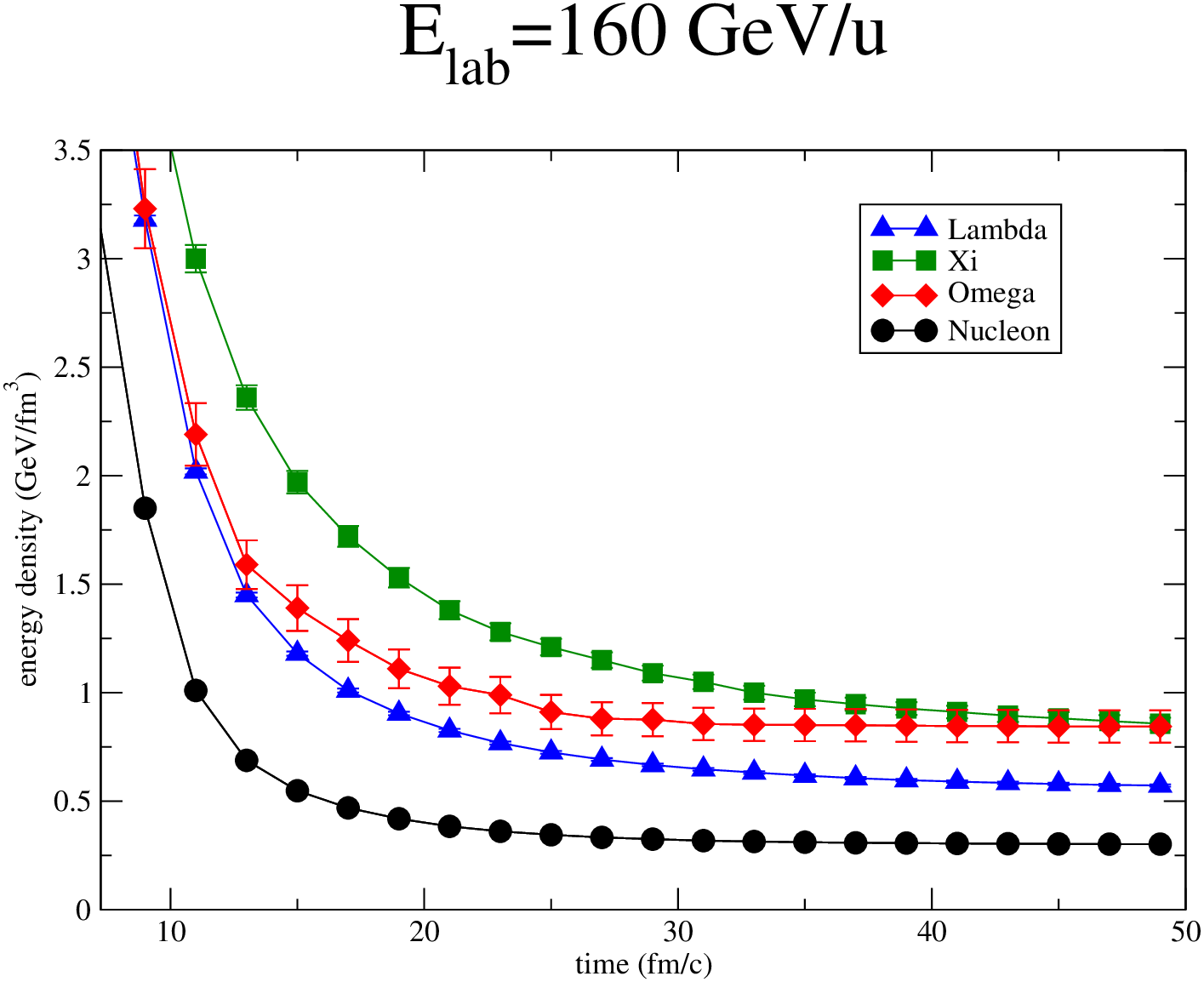}
 \caption{The average freeze-out energy density for Nucleons, Lambda's, Xi's and Omega's as a function
 of time for $E_{lab}$=20GeV/u (left) and $E_{lab}$=160GeV/u (right). A saturation
 of the respective curve indicates that the majority of particles of that species have ceased
 interacting.}
\end{figure}
\end{center}

If the picture we have developed regarding the dynamics of hadron
freeze-out and its flavor sensitivities is correct, then a nucleon
and a $\Omega$ baryon freezing out {\it at the same time} within
similar conditions (i.e. mid-rapidity) should
be probing the same medium in terms of its temperature and energy-density:

\begin{center}
\begin{figure}
\includegraphics[width=0.47\textwidth, angle=0]{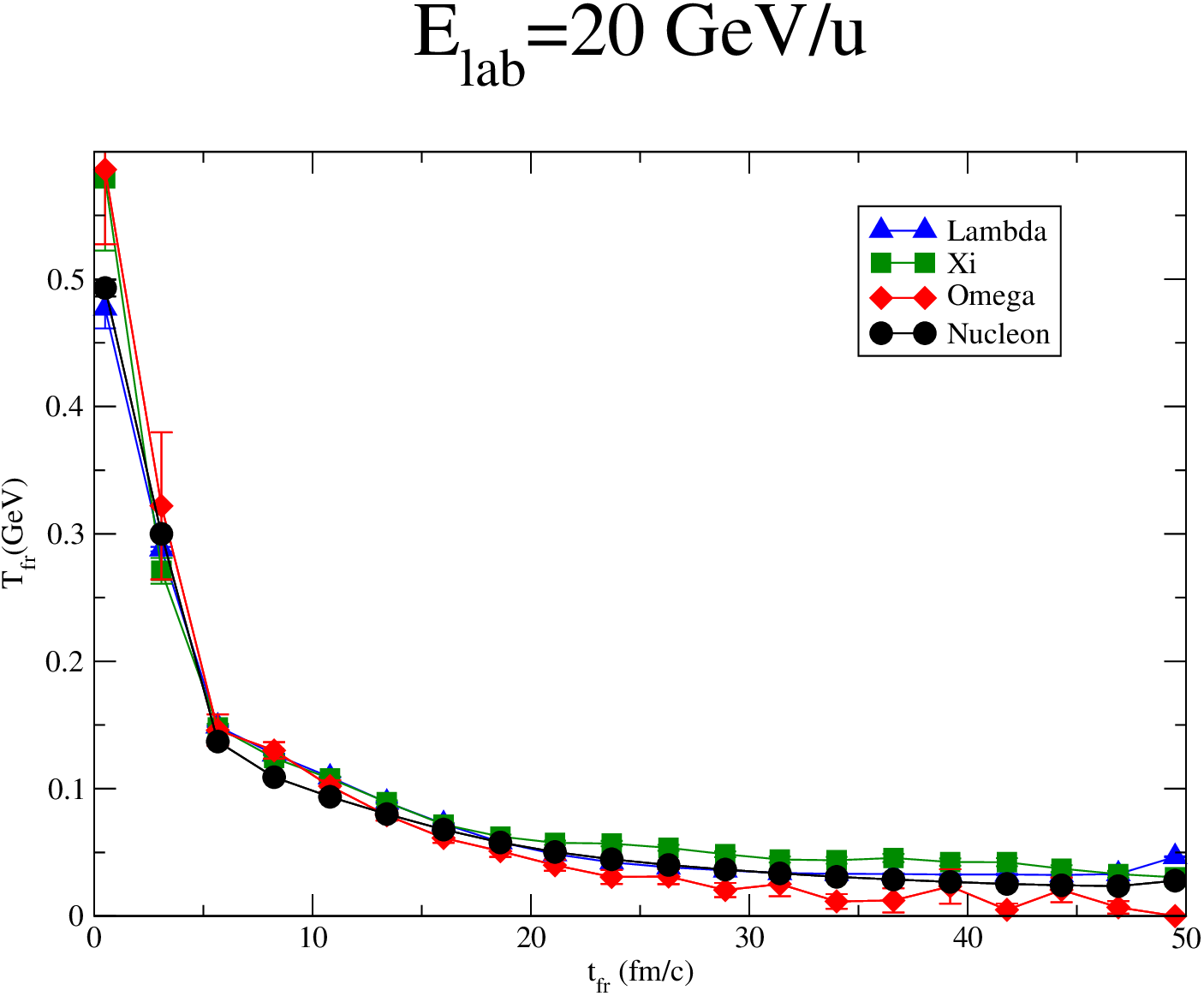}\hfill
\includegraphics[width=0.47\textwidth, angle=0]{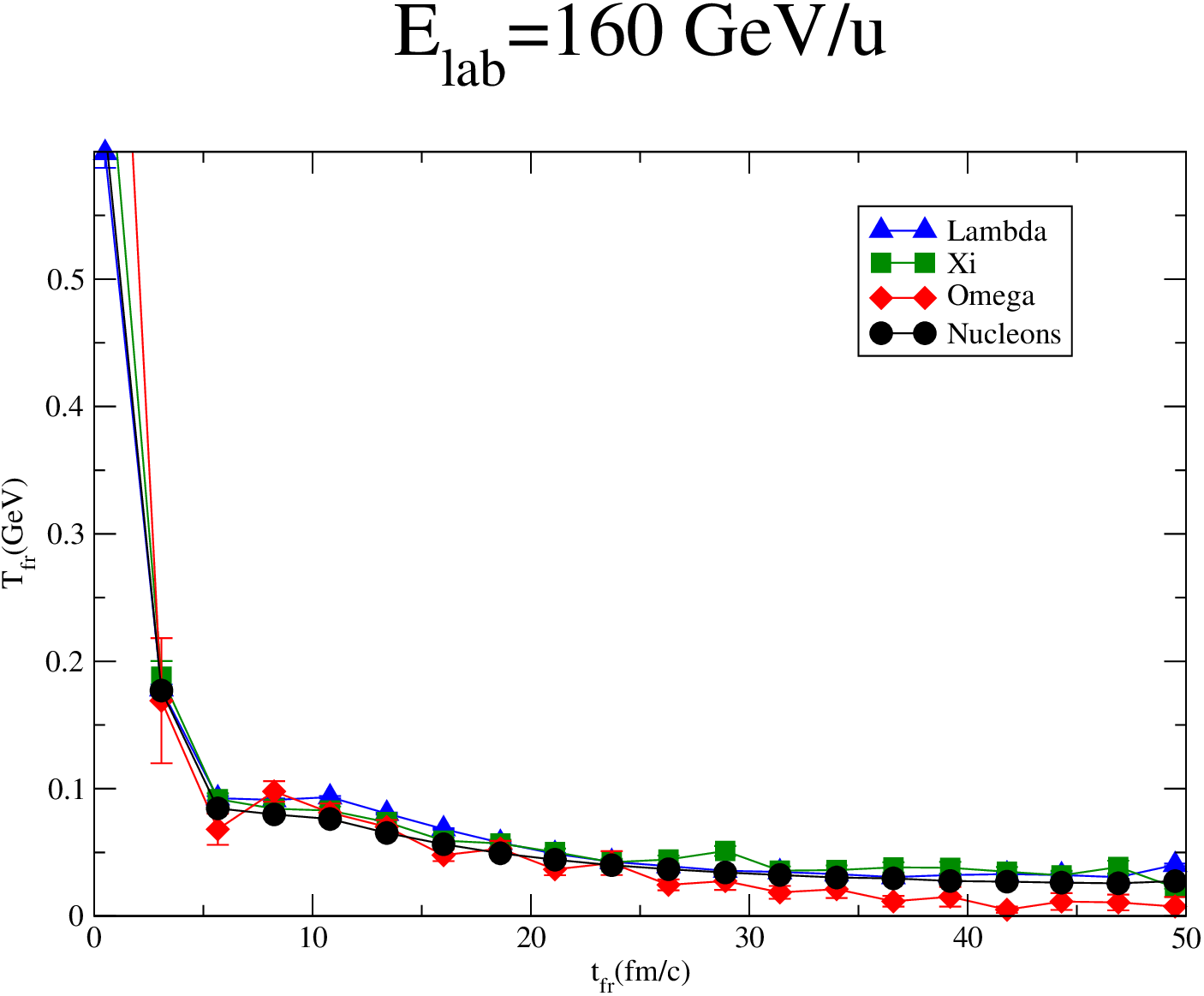}
 \caption{Freeze-out temperature as a function of freeze-out time for Nucleons, 
 Lambda's, Xi's and Omega's  for $E_{lab}$=20GeV/u (left) and $E_{lab}$=160GeV/u (right).
 The agreement between the curves indicates that baryons freezing out at identical times
 probe the same medium, irrespective of their species.}
\end{figure}
\end{center}

Figure~5 shows the average freeze-out temperature for all particles freezing out
{\em at a given time}, i.e. for each time-step we evaluate the freeze-out temperature
only for those baryons whose final interaction actually occurs in that timestep.
The figures (left for 20 GeV/nucleon and right for 160 GeV/nucleon incident beam energy)
confirm our hypothesis -- all four baryon species exhibit near identical freeze-out
temperatures for final interactions occurring at identical times.

Let us now investigate how our results depend on the incident beam energy
of the heavy-ion collision: Figure~6 shows the excitation function of freeze-out
temperatures for the four different baryon species vs. incident beam energy for central 
Au+Au collisions. The left frame displays the average freeze-out temperature at the average 
freeze-out time for each species, whereas the right frame takes the freeze-out temperature
at the time when 80\% of all particles of each species have ceased interacting. Naturally
in the latter case we obtain somewhat lower freeze-out temperatures due to later freeze-out
times (and thus lower freeze-out temperatures) contributing to the analysis. The only other
quantitative difference between the two analyses is the $\Xi$, since for the average freeze-out
time criterion we encounter contamination by the long-lived $\Xi^*(1530)$ resonance, leading to a 
larger average freeze-out time for the $\Xi$.

\begin{center}
\begin{figure}
\includegraphics[width=0.47\textwidth, angle=0]{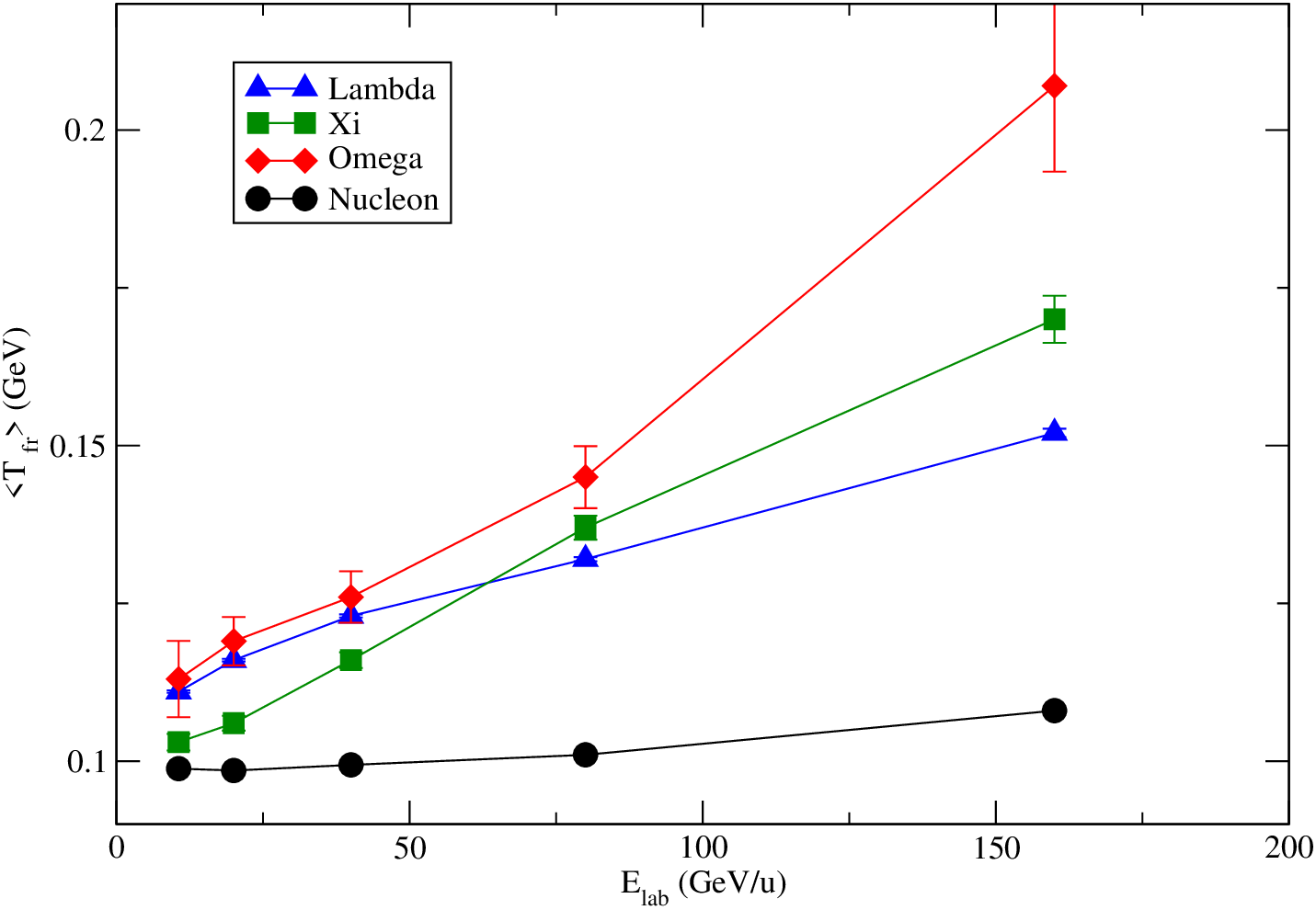}\hfill
\includegraphics[width=0.47\textwidth, angle=0]{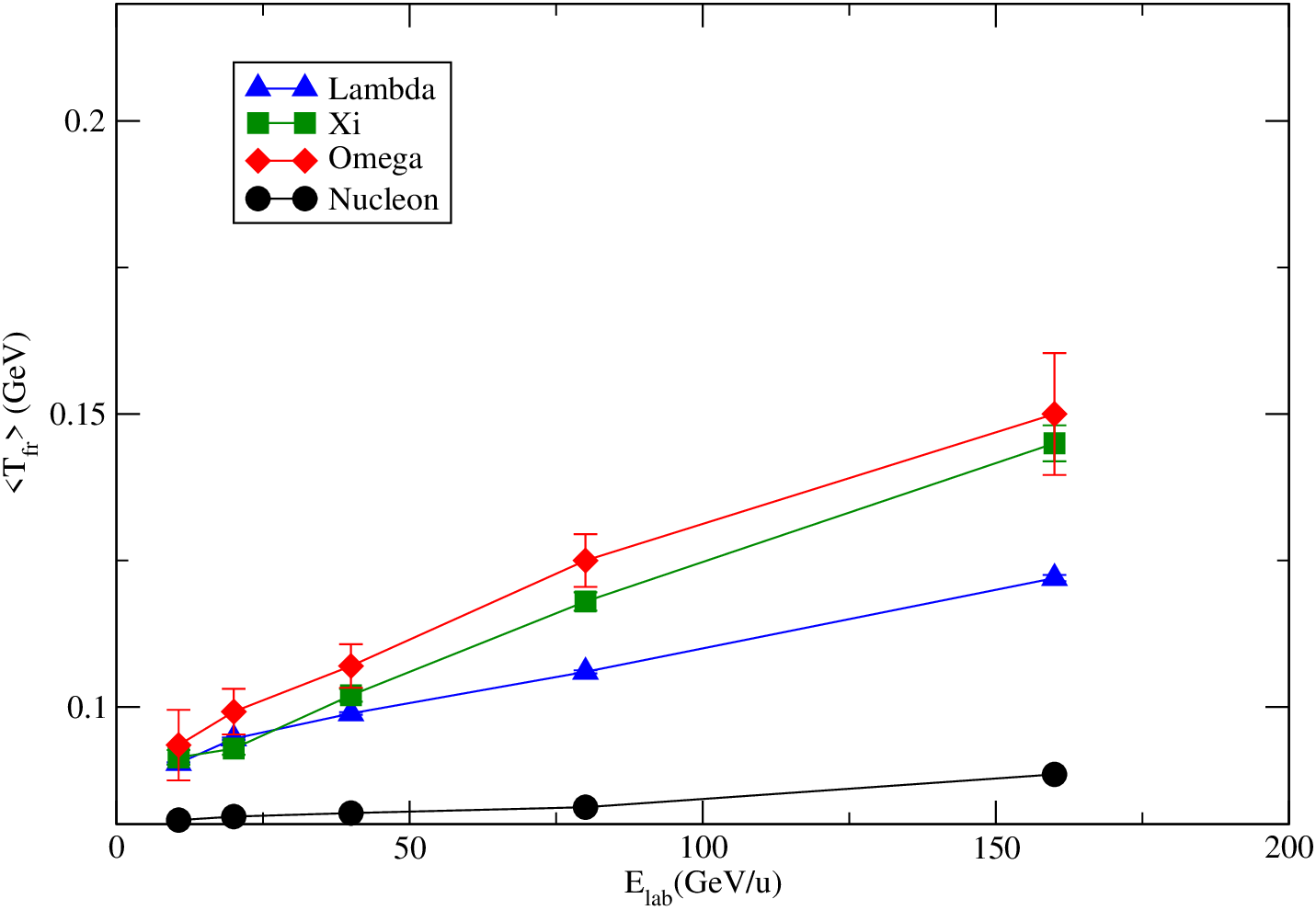}
 \caption{Average freeze-out temperature vs beam energy for final state baryons. The freeze-out
 temperature is calculated at the average freeze-out time (left) and at the time when 80 percent of  Nucleons, Lambda, Xi and Omega have ceased interacting (right).}
\end{figure}
\end{center}

Figure~7 shows the same analysis as Figure~6, but with the freeze-out energy-density
instead of freeze-out temperature. It is interesting to note that for both quantities
and the more realistic 80\% criterion, at 160 GeV/nucleon incident beam energy 
the $\Omega$ exhibits a freeze-out temperature and energy-density very close to the
critical temperature and energy-density of the deconfinement phase transition.

\begin{center}
\begin{figure}
\includegraphics[width=0.47\textwidth, angle=0]{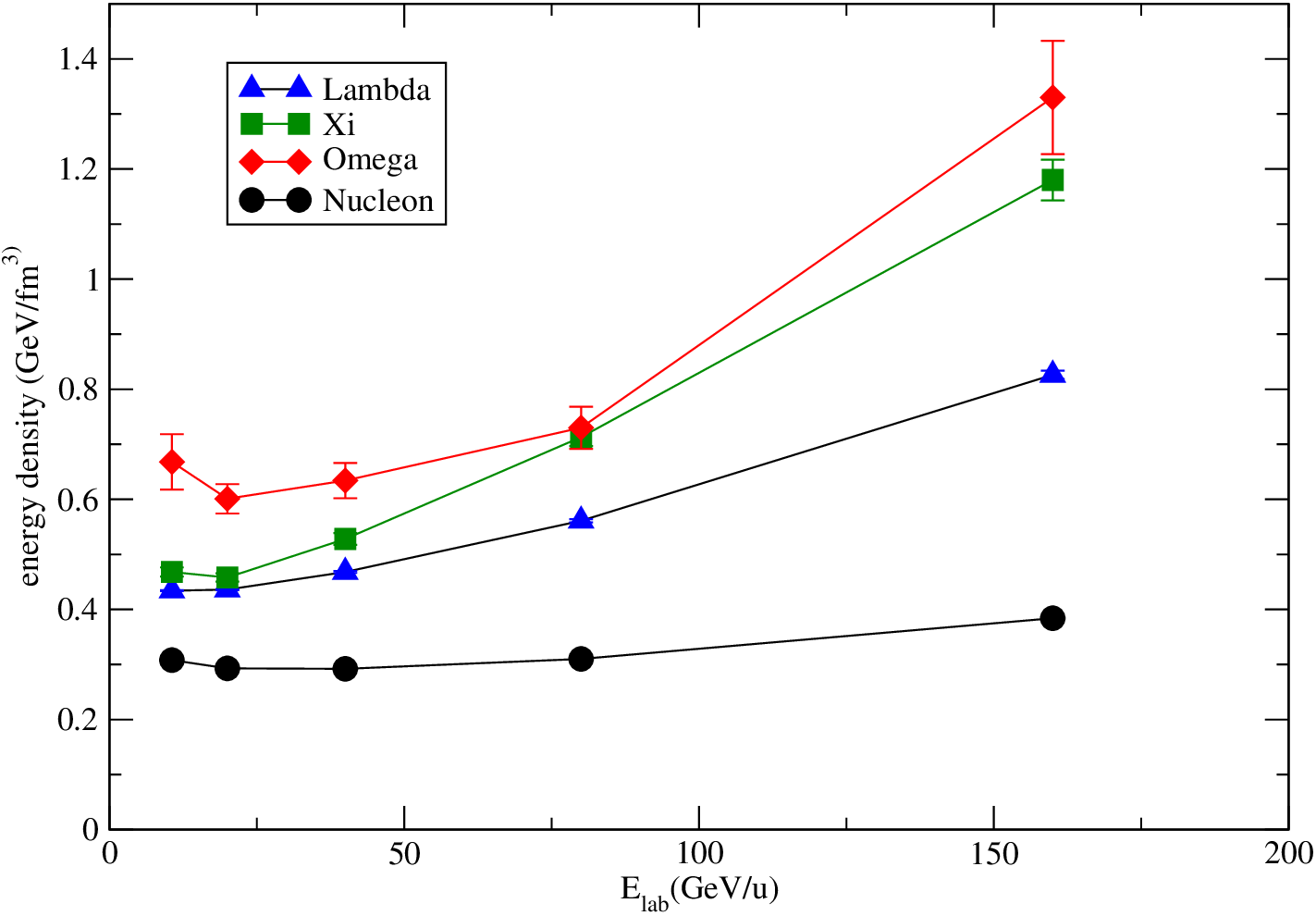}\hfill
\includegraphics[width=0.47\textwidth, angle=0]{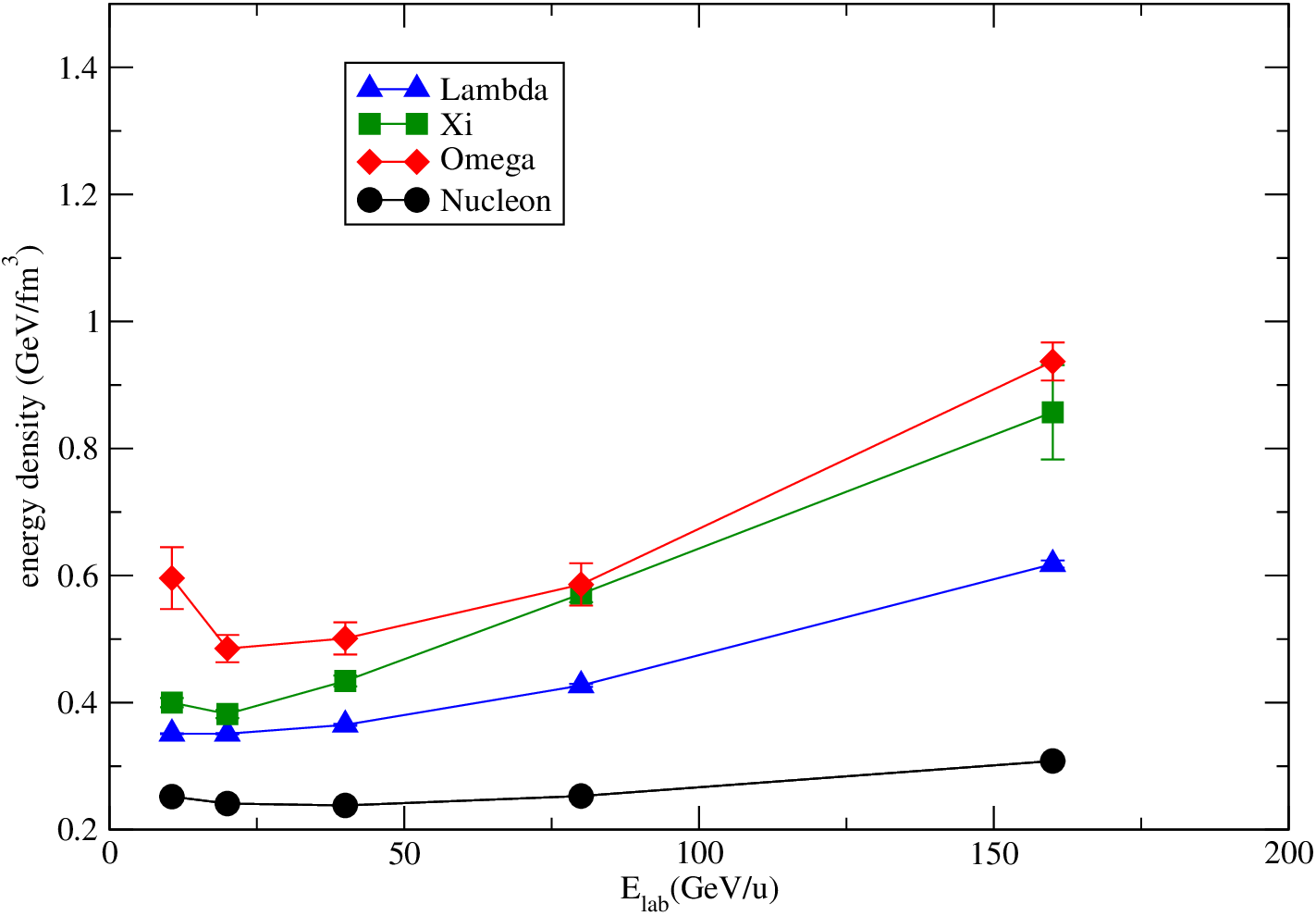}
 \caption{Aaverage freeze-out energy density vs beam energy for final state baryons. The freeze-out
 energy-density is calculated at the average freeze-out time (left) and at the time when 80 percent of  Nucleons, Lambda, Xi and Omega have ceased interacting (right).}
\end{figure}
\end{center}

\section{Summary}
We have calculated the excitation function of freeze-out time, average freeze-out temperature 
and freeze-out energy-density of final state baryons created in relativistic heavy-ion collisions
within the UrQMD model. We find that the
$\Omega$ on average freezes out
earlier than the nucleon, $\Lambda$ and $\Xi$ and thus exhibits a higher freeze-out temperature
and energy-density. In general, baryons systematically freeze out earlier with rising strangeness
content. The average freeze-out temperature and energy
density as well as the spread between the different baryonic states
increase monotonously with increasing beam energy and approaches the values
of the critical temperature and energy density of the QCD deconfinement transition at
an incident beam energy of 160 GeV/nucleon. Our findings give rise to the speculation
that for the $\Omega$ this should be a universal value in the case of a 
hadronizing Quark-Gluon-Plasma.

\section{Acknowledgments}
This work was supported by the DOE under grant DE-FG02-05ER41367 and
the CSC Scholarship programm. Zhenglian Xie would like to thank
Nankai University for giving the chance to do the research work in
Duke University. Zhenglian Xie also thanks Prof. Berndt Mueller for
helpful discussions and Duke University for the hospitality during
her one year stay for the research work.

\section*{References}
\bibliographystyle{iopart-num}
\bibliography{/Users/bass/Publications/SABrefs}
\end{document}